\documentclass[twocolumn,showpacs,preprintnumbers,amsmath,amssymb]{revtex4-1}

\usepackage{graphicx}% Include figure files
\usepackage{dcolumn}% Align table columns on decimal point
\usepackage{bm}% bold math

\begin{document}

%\preprint{??????}

\title{Two-photon production of dilepton pairs in peripheral heavy ion collisions}
\author{Spencer R. Klein} \affiliation{Lawrence Berkeley National Laboratory, Berkeley CA, 94720, USA}
 
\date{\today}

\begin{abstract}

The STAR collaboration has observed an excess production of $e^+e^-$ pairs in relativistic heavy ion collisions, over the expectations from hadronic production models.  The excess pairs have transverse momenta $p_T < 150\ {\rm MeV}/c$ and are most prominent in peripheral gold-gold and uranium-uranium collisions.  The pairs exhibit a peak at the $J/\psi$ mass, but include a wide continuum, with pair invariant masses from 400 MeV/c$^2$ up to 2.6 GeV/c$^2$.  The ALICE Collaboration observes a similar excess in peripheral lead-lead collisions, but only at the $J/\psi$ mass, without a corresponding continuum.

This paper presents a calculation of the cross-section and kinematic for two-photon production of $e^+e^-$ pairs, and find general agreement with the STAR data.  The calculation is based on the STARlight simulation code, which is based on the Weizs\"acker-Williams virtual photon approach.   The STAR continuum observations are compatible with two-photon production of $e^+e^-$ pairs.   The ALICE analysis required individual muon $p_T$ be greater than 1 GeV/c; this eliminated almost all of the pairs from two-photon interactions, while leaving most of the $J/\psi$ decays.

\end{abstract}

\maketitle

\section{Introduction}

Two-photon collisions were extensively studied at $e^+e^-$ colliders, where each lepton emitted a photon.  These reactions were used to study a variety of hadronic final states and also four-lepton final states \cite{Budnev:1974de,who}.  More recently, they have been studied in ultra-peripheral collisions (UPCs) of relativistic heavy ions.   UPCs are collisions where the nuclei physically miss each other (impact parameter $b$ greater than twice the nuclear radius $R_A$), but interact electromagnetically.   The very strong electromagnetic fields which emanate from highly charged heavy ions lead to large cross-sections for photonuclear and two-photon interactions \cite{Baur:2001jj,Bertulani:2005ru,Baltz:2007kq}.   Two-photon interactions of interest include $e^+e^-$ production in strong fields and light-by-light scattering \cite{Aaboud:2017bwk}.   
The reaction $\gamma\gamma\rightarrow l^+l^-$ has been studied in UPCs by the STAR \cite{Adams:2004rz}, ATLAS \cite{Dyndal:2017wbv}, ALICE \cite{Abbas:2013oua} and CMS \cite{Khachatryan:2016qhq} collaborations, and good agreement with lowest order quantum electrodynamics predictions was seen.

Recently, the STAR experiment observed an excess of $e^+e^-$ pairs, produced at small transverse momentum ($p_T < $ 150 MeV/c) in peripheral gold-gold and uranium-uranium collisions at a center of mass energies of 200 GeV/nucleon pair  and 193 GeV/nucleon pair respectively \cite{Shuai,Wangmei}.   The $e^+e^-$ invariant mass spectrum of the low $p_T$ excess shows a significant peak at the $J/\psi$, but also includes a continuum component in the mass range from 400 MeV to 2.6 GeV.  
The signal is most prominent in more peripheral collisions (60\% to 80\% centrality) than in those with smaller impact parameters.   Here, 0\% centrality is a head-on collision with impact parameter $b=0$, while 100\% centrality is a grazing collision with $b=2R_A$.  

ALICE has also reported on an excess of $\mu^+\mu^-$ pairs at low $p_T$ \cite{Adam:2015gba} in very peripheral (70-90\% centrality) lead-lead collisions at  a center of mass energy of 2.76 TeV/nucleon pair.  The peak data are consistent with $J/\psi$ photoproduction  \cite{Klusek-Gawenda:2015hja,Zha:2017jch,Shi:2017qep}, but they do not see significant continuum production.

The  STAR and ALICE $J/\psi$ rates and $p_T$ spectra are in agreement with expectations from photoproduction, including the cutoff at very low $p_T$ due to interference between photoproduction for photons coming from the opposite directions \cite{Klein:1999gv,Abelev:2008ew}.   However, the broad STAR mass continuum does not seem compatible with vector meson photoproduction.  The featureless mass distribution and limitation to low $p_T$ are both very suggestive of two-photon production of $e^+e^-$ pairs.

In this work, we calculate the rates and kinematic distributions for two-photon production of $e^+e^-$ pairs in peripheral hadronic collisions, and show that it is generally compatible with the STAR observation and ALICE non-observation of continuum production at low $p_T$.   The results provide a basis for a more detailed comparison between the data and two-photon theory.

\section{Methods}

The cross-sections and kinematic distributions for $\gamma\gamma\rightarrow e^+e^-$ in peripheral collisions are calculated using the photon flux predicted by the Weizs\"acker-Williams method, and the lowest order Breit-Wheeler cross-section for $\gamma\gamma\rightarrow e^+e^-$.   For ultra-relativistic particles, the photon flux at a perpendicular distance $b$ from an emitting nucleus with nuclear charge $Z$ is \cite{Baur:2001jj,Bertulani:2005ru,Baltz:2009jk}
\begin{equation}
N(k,b) = \frac{Z^2\alpha}{\pi^2} \frac{k}{(\hbar c\gamma)^2} K_1(x)^2
\label{eq:flux}
\end{equation}
where $k$ is the photon energy, $x=kb/\gamma\hbar c$, $\gamma$ is the ion Lorentz boost, $\alpha\approx 1/137$ is the electromagnetic fine structure constant, and $K_1(x)$ is a modified Bessel function. 

The cross-section depends on the overlap of the photon fluxes, integrated over all possible transverse positions for the two ions and the location of the photon-photon interaction.  This can be simplified to a three-dimensional integral over the distance from the first ion to the interaction site, $b_1$, the distance from the second ion site, $b_2$, and the angle $\phi$ between the two ion-interaction site vectors \cite{Baur:1990fx}.  For peripheral collisions, the ion-ion impact parameter range is restricted to match a desired centrality bin: $b_{\rm min} < |b| < b_{\rm max}$.    The cross-section to produce a final state $W$ from photons with energy $k_1$ and $k_2$ is 
\begin{align}
\sigma = \int_{R_A}^\infty \pi b_1d^2 b_1 \int_{R_A}^\infty \pi b_2 d b_2 \int_0^{2\pi} d\phi N(k_1,b_1) N(k_2,b_2) \\ \sigma(k_1k_2\rightarrow W) \theta(b_{\rm min} < |b_1-b_2| < b_{\rm max}) \nonumber
\label{eq:sigma}
\end{align}
where the $\theta$ function is 1 when the inequality is satisfied, and 0 otherwise.

This approach assumes that the photons come from the electromagnetic fields of the entire nucleus, moving at the full beam velocity.   Since the fields at time $t$ are evaluated based on the nucleus configuration at a retarded time $\tau=t-|b_i|/\gamma c$ \cite{Jackson}, this assumption should be satisfied.  This approach also assumes that the photon flux is zero within the emitting nucleus ({\it i. e.} for $|b|<R_A$).  The inclusion of interactions occurring within one of the nuclei would slightly increase the calculated cross-section, with the size of the increase rising with increasing pair mass.

The final state pair mass $M_{ll}$ is given by the two-photon center of mass energy $W$.  The photon energies map into $W$ and rapidity $y$ via $W=M_{ll}=2\sqrt{k_1k_2}$ and $y= 1/2 \ln{(k_1/k_2)}$.   

The cross-section to produce pairs of leptons with lepton mass $m$ is the Breit-Wheeler cross-section \cite{Brodsky:1971ud}:
\begin{align}
\nonumber 
\sigma(\gamma\gamma\rightarrow l^+l^-) = \\ 
\frac{4\pi\alpha^2}{W^2} 
\bigg[\big(2+\frac{8m^2}{W^2} - \frac{16m^4}{W^4}\big) \ln{(\frac{W+\sqrt{W^2-4m^2}}{2m})} \\
\nonumber -\sqrt{1-\frac{4m^2}{W^2}}\big(1+\frac{4m^2}{W^2}\big)\bigg].
\label{eq:breitwheeler}
\end{align}
The angular distribution of the decay electrons also follows Breit-Wheeler, with the leptons preferentially emitted in the forward and backward directions:
\begin{equation}
G(\theta) = 2 + 4\big(1-\frac{4m^2}{W^2}\big)
\frac{(1-\frac{4m^2}{W^2})\sin^2(\theta)\cos^2(\theta)+ \frac{4m^2}{W^2}}
{(1-(1-\frac{4m^2}{W^2})\cos^2(\theta))^2}.
\label{eq:angle}
\end{equation}
where $\theta$ is the angle between the beam direction and one of the leptons, in the lepton-lepton center of mass frame.  

The pair $p_T$ is the vector sum of the photon $k_T$; the photon $k_T$ comes from the Weizs\"acker-Williams approach \cite{Vidovic:1992ik,Klein:1999gv}:
\begin{equation}
\frac{dN}{dk_T} = \frac{2F^2(k^2/\gamma^2 + k_T^2)k_T^3}{(2\pi)^2((k/\gamma)^2 + k_T^2)^2}
\label{eq:photonpt}
\end{equation}
where $F$ is the nucleon form factor, per Ref. \cite{Klein:2016yzr}.  The individual lepton $p_T$ includes contributions from this initial $\gamma\gamma$ $p_T$, plus the transverse kick acquired from the non-zero $\theta$ in Eq. \ref{eq:angle}.

The calculations are done in the framework of the STARlight Monte Carlo \cite{Baltz:2009jk,Klein:2016yzr}.   We  modified STARlight to limit the range of integration in impact parameter to a user selectable range, regardless of whether the two nuclei overlap or not, as shown in Eq. 2; this code is publicly available in the trunk of STARlight \cite{STARlight}.   STARlight has been extensively compared with UPC data, with good agreement found for $\gamma\gamma\rightarrow l^+l^-$, with data from STAR \cite{Adams:2004rz}, ATLAS \cite{Dyndal:2017wbv}, ALICE \cite{Abbas:2013oua} and CMS \cite{Khachatryan:2016qhq} collaborations.  There is a discrepancy at small pair $p_T$ where the equivalent photon approximation predicts an overabundance of pairs, compared to both a lowest order Quantum Electrodynamics (QED) calculation and data \cite{Adams:2004rz}.  ATLAS also sees a small tail of events with larger pair $p_T$; the collaboration notes that this might be background, or it might be from the two-photon signal.  Also, STARlight assumes that nuclei are spherical.  Uranium is aspherical; this introduces an additional uncertainty in the uranium-uranium calculations.

\section{Results}

Most of the $\gamma\gamma\rightarrow ee$ cross-section is for near-threshold pairs, which are not visible in existing detectors.  These calculations focus on experimentally accessible interactions, so consider only pairs with invariant masses above 400 MeV/c$^2$.    Results are presented for five different experimental conditions: 60-80\% centrality, 40-60\% centrality and 10-40\% centrality Au-Au collisions at a center-of-mass (CM) energy of 200 GeV/nucleon, 60-80\% U-U collisions at a slightly lower CM energy, 193 GeV/nucleon (all at RHIC), and 60-80\% Pb-Pb collisions at a center of mass energy of 2.76 TeV/nucleon at the LHC.  These calculations are for a central detector, following the STAR acceptance \cite{Shuai}.  As noted below, the ALICE forward muon spectrometer has little acceptance for two-photon production of dimuons, and an ALICE study in the central region would likely have a similar acceptance to STAR. 

\begin{table*}
\begin{tabular}{lrrrrrr}
Ion/ Centrality & $b-$ range & \ \ \ \ $\sigma_{\rm had}$ &\ \ \  $\sigma_{\rm ll}$ (restr.) & \% satisfying     & $\#_{\rm visible\ ll}$ \\
                                &                   &                                        &                                               & $l^\pm$ criteria &\ \ \  hadronic event  \\
\hline
60-80\% RHIC Au-Au\ \ \     & \ \ 11.4-13.2 fm  & 1.36 b & 3.7 mb & 3.3\% & $8.9\times10^{-5}$  &  \\
40-60\% RHIC Au-Au    & 9.4-11.6 fm   & 1.36 b & 3.8 mb   & 3.3\% & $9.2\times10^{-5}$ & \\
10-40\% RHIC Au-Au    & 4.8-9.4 fm   & 2.04 b & 6.4 mb   &  3.3\% & $1.03\times10^{-4}$ &  \\
\hline
60-80\% RHIC U-U        &14.1-15.8 fm & 1.56 b & 5.2 mb & 3.3\%  &  $1.01\times10^{-4}$ & \\
\hline
60-80\% LHC  Pb-Pb     &13.0-14.7 fm & 1.51 b & 6.4 mb   & 3.5\%     & $1.4\times10^{-4}$ & \\
\hline
\end{tabular}
\caption{Ions and centralities, impact parameters and cross-sections for two-photon production of lepton pairs.  The centralities were chosen to match the STAR analysis.  
The table also gives the calculated photoproduction cross-section (within the given constraints on pair mass and rapidity), the fraction of those events that pass the individual lepton kinematic cuts, and the fraction of the hadronic events in that centrality that should contain a visible (satisfying the pair and individual lepton constraints) lepton pair. 
\label{table:sigma}}
\end{table*}

The collaborations report impact parameter range in terms of collision centrality.  A Monte Carlo Glauber calculation \cite{Loizides:2017ack} is used to convert from the reported centralities into impact parameters.  The calculation finds cross-sections of 6.8 barns for Au-Au collisions at RHIC, 7.8 barns for U-U collisions at RHIC \cite{DavidD}, and 7.6 barns for Pb-Pb collisions at the LHC at a center of mass energy of 2.76 TeV/nucleon pair, all with errors that are negligible compared to other uncertainties in the overall calculation.  These cross-sections are 4-8\% lower than some other results \cite{Fischer:2013uwj},  likely because the calculation uses slightly lower inelastic proton-proton cross-sections than other works.   We then use a simple black-disk model to convert from centrality to impact parameter range,  so 100\% centrality corresponds to $b_{\rm max} = \sqrt{\sigma_{\rm had}}/\pi$.   Table \ref{table:sigma} shows the centrality regions and hadronic cross-sections for those centralities.

\begin{figure}[tb]
  \includegraphics[width=0.46\textwidth]{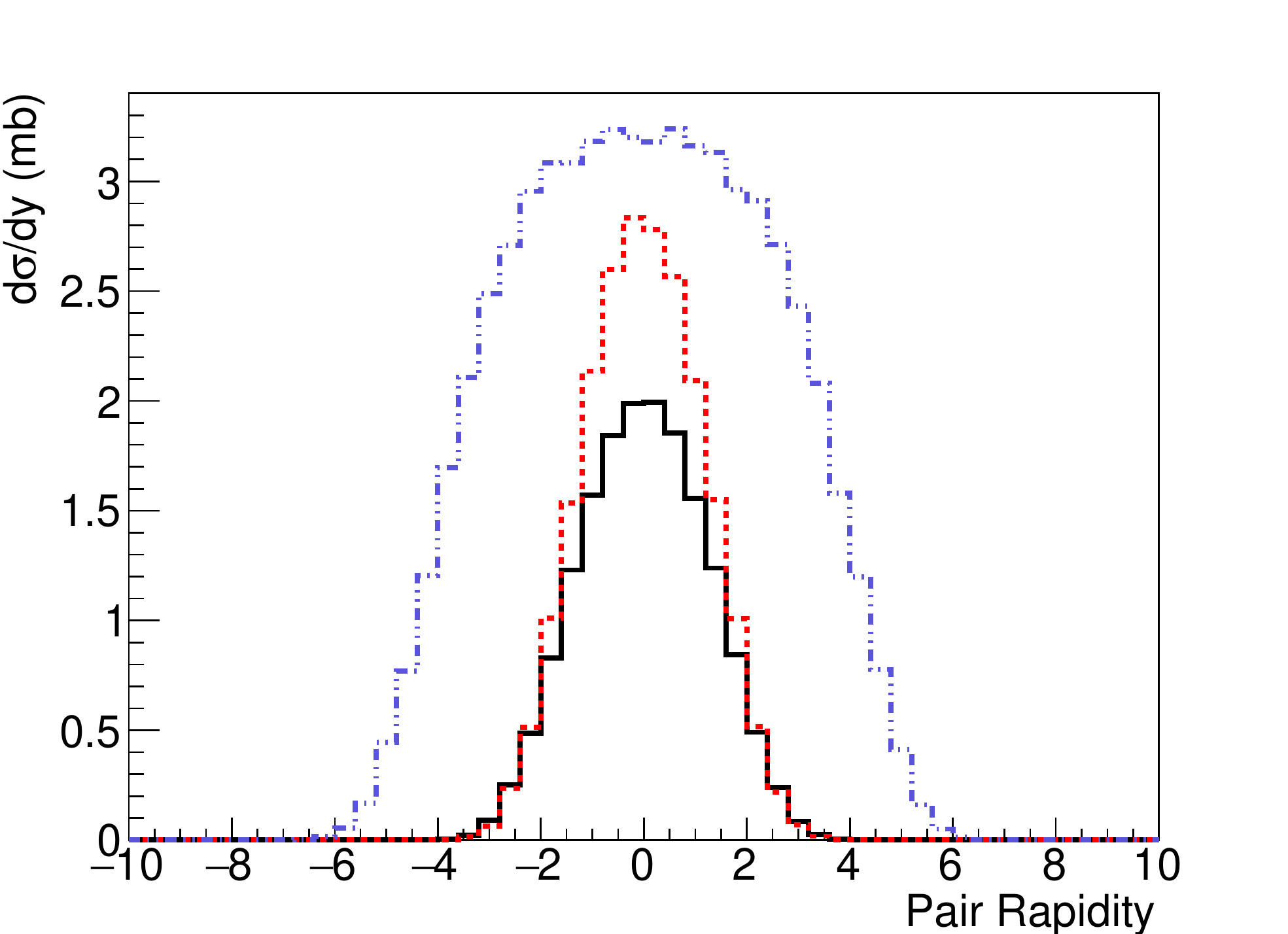}
\caption{$d\sigma/dy$ for $\gamma\gamma\rightarrow e^+e^-$ with pair mass more than 400 MeV/c$^2$, for 60-80\% centrality gold-gold (solid black histogram) and uranium-uranium (dashed red histogram) at RHIC and lead-lead collisions at the LHC (dot-dashed blue line).
\label{fig:dsdy}
}
\end{figure}

Figure \ref{fig:dsdy} shows the rapidity distribution $d\sigma/dy$ for 60-80\% centrality collisions of gold and uranium at RHIC and lead at the LHC for $M_{ee} > 0.4$ GeV/c$^2$.  Because the distribution is almost independent of centrality, only one RHIC curve is shown.  The integrated cross-sections are  6.7 mb, 8.7 mb and 24.2 mb for gold, uranium, and lead respectively.    The LHC cross-section is much larger and covers a much wider rapidity range than the RHIC curves, because of the higher beam energy.   The uranium cross-section is about 36\% larger than that for gold, less than the 54\% increase expected from the naive $Z^4$ scaling.   Uranium nuclei are larger than gold nuclei, and the per-nucleon collision energy was a bit lower , so the photon flux, Eq. \ref{eq:flux} is cut off at about 10\% lower energy than for gold.  The cutoff is  also evident in the rapidity distribution, which is slightly narrower than for gold-gold. 

\begin{figure}[tb]
 \includegraphics[width=0.46\textwidth]{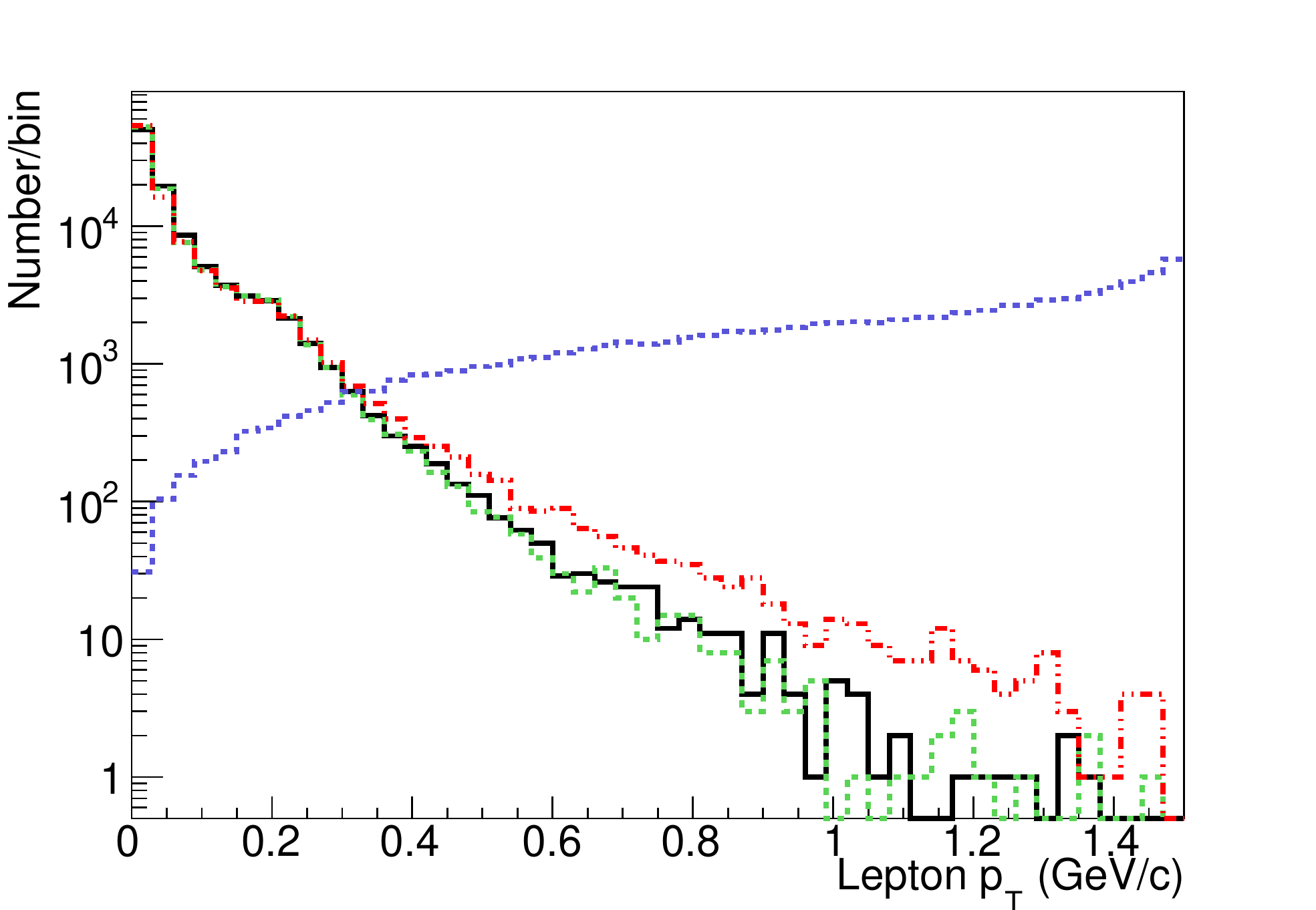}
\caption{Individual lepton $p_T$ for Au-Au (solid black histogram) and U-U (dashed green histogram) at RHIC and Pb-Pb collisions at the LHC (dot-dashed red histogram), along with the lepton $p_T$ from photoproduction of $J/\psi$ in Au-Au ultra-peripheral collisions at RHIC (dotted blue line).
\label{fig:leptonpt}
}
\end{figure}

Figure \ref{fig:leptonpt} shows the $p_T$ spectra for the individual leptons for the three systems, along with, for comparison, the lepton $p_T$ from photoproduction of $J/\psi$ in gold-gold ultra-peripheral collisions at RHIC \cite{Klein:1999qj,Klein:1999gv,Klein:2016yzr}.  The leptons from $\gamma\gamma\rightarrow e^+e^-$ are peaked at very low $p_T$,in sharp contrast to the leptons from $J/\psi$ decays.  This difference immediately shows why the ALICE forward muon spectrometer cut on muon $p_T > 1$ GeV/c almost completely eliminates  pairs from $\gamma\gamma$ interactions while retaining the pairs from coherent $J/\psi$ photoproduction, even though the two  reactions have not dissimilar pair $p_T$ spectra.    The $p_T$ spectra from the three $\gamma\gamma$ channels are similar, with small changes due to the per-nucleon collision energy and size of the nuclei.  

We now turn to calculations of cross-sections within the STAR acceptance: pairs with pair mass $M_{ee} > 0.4$ GeV/c$^2$ and rapidity $|y_{ee}|<1$.   STAR also requires that the individual leptons satisfy $p_{T,e}> 200 $ MeV/c and pseudorapidity $|\eta_e|<1$.     
Table \ref{table:sigma} shows the cross-sections for five different beam/energy/centrality conditions, for hadronic interactions, the cross-sections for $\gamma\gamma\rightarrow e^+e^-$ within the pair rapidity and pair mass range, the probability for thosepaper events to also satisfy the individual lepton pseudorapidity and $p_T$ cuts, and, finally, the number of pairs within the full STAR acceptance per hadronic collisions.    The hadronic cross-section is the total hadronic cross-section times the width of the centrality bin.  

For a fixed condition and acceptance, the restricted two-photon cross-section depends mostly on the width of the centrality bin.  The cross-section is higher for Pb-Pb collisions, because of the higher LHC collision energy; the increase in $\gamma\gamma$ cross-section with energy is much faster than the rise in hadronic cross-section.  This increase is reflected in the higher number of visible $ee$ pairs for Pb-Pb collisions than for the lower energy RHIC systems. 

The restricted cross-section is 40\% larger for U-U cross-sections than for Au-Au collisions.  This is again less than the 54\% increase expected from the $Z^4$ scaling, but larger than the increase in the all-rapidity cross-section, because, for the heavier nucleus, production is more concentrated at small $|y|$.   

The fraction of lepton pairs that pass the individual lepton cuts is about 3.4\%, almost independent of the collision conditions, with only a small rise for the higher-energy Pb-Pb collisions.  This acceptance is so low because, per Eq. 4, the pairs from two-photon interactions prefer a forward-backward geometry, and so avoid the central region.

 \begin{figure}
 \includegraphics[width=0.5\textwidth]{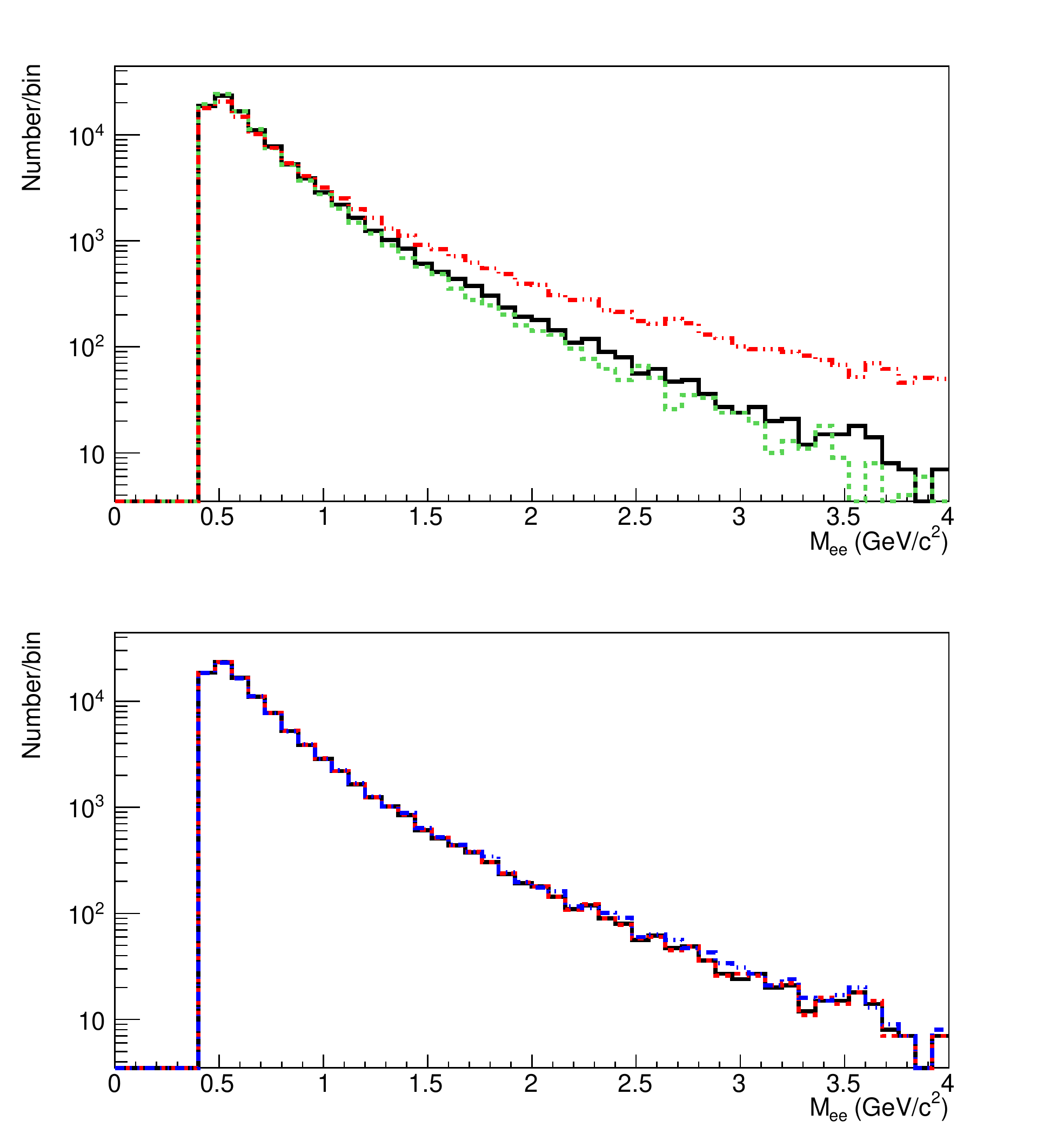}
\caption{Pair invariant mass spectra for (top) Au-Au (black solid histogram), U-U (green dashed histogram), and Pb-Pb (red dot-dashed histogram) and (bottom) 60-80\% centrality (black solid histogram), 40-60 \% centrality (red dashed histogram), and 10-40\% centrality (blue dot-dashed histogram).  The bottom three histograms are  indistinguishable.  
\label{fig:leptonmass}
}
\end{figure}

Figure \ref{fig:leptonmass} shows the pair invariant mass distributions for events within the STAR acceptance.   The Pb-Pb data spectrum is harder than the Au-Au and U-U distributions, because of the higher beam energy.  The U-U spectrum is slightly softer than the Au-Au distribution, because of the slightly larger nuclear size and lower per-nucleon collision energy.  The shape of these distributions are similar to the STAR data presented in Ref. \cite{Shuai}; The number of events drops by roughly a factor of 10 as the pair mass doubles from 0.5 GeV to 1.0 GeV, in at least rough agreement with the STAR data.  

 \begin{figure}
 \includegraphics[width=0.5\textwidth]{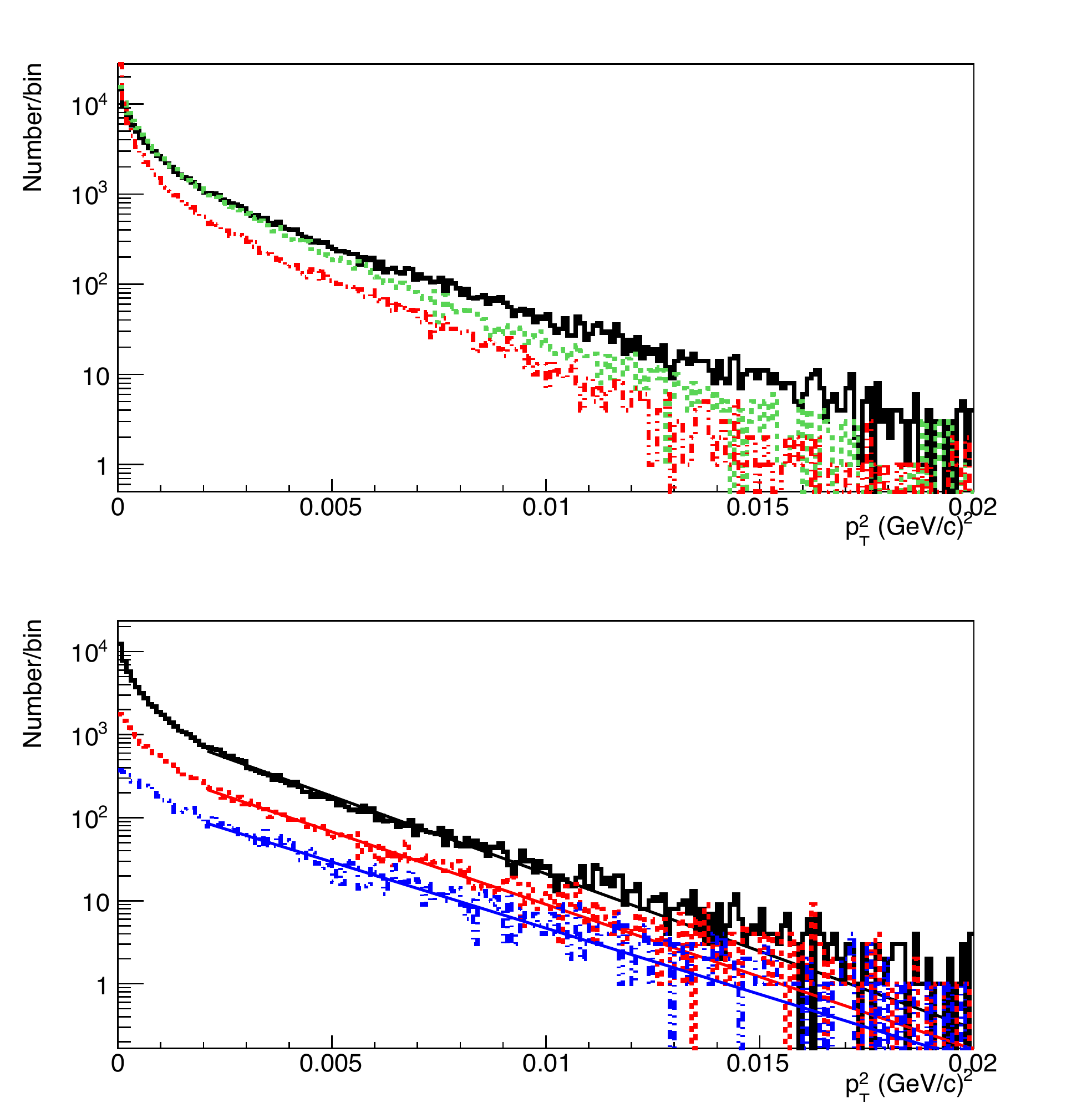}
\caption{Pair $p_T^2$ spectra for (top) Au-Au (black solid histogram), U-U (green dashed histogram), and Pb-Pb (red dot-dashed histogram) and (bottom) for three invariant mass ranges: 0.4 to 0.76 GeV/c$^2$ (black solid histogram), 0.76-1.2 GeV/c$^2$ (red dashed histogram), and  1.2-2.6 GeV/c$^2$ (blue dot-dashed histogram).  The bottom three histograms show a clear mass ordering.  The lines in the bottom plot are fits to Eq. 6 in the displayed region, as discussed in the text.
\label{fig:pairpt}
}
\end{figure}

Figure \ref{fig:pairpt} shows the distribution of lepton-pair $p_T^2$ for the three species (top), and the three Au-Au centralities (bottom).   A significant upturn is seen for $p_T^2 < 0.002$ (GeV/c)$^2$, while at higher energies, the distribution looks quasi-exponential.  The $p_T^2$ scale is lowest for Pb-Pb because, for a fixed photon energy $k$, the photon $k_T$ drops with increasing ion energy.  The U-U distribution is slightly softer than the Au-Au spectrum because the larger nuclear size softens the energy distribution. 

At low $p_T$, the equivalent photon approach used here differs from a lowest-order QED calculation, which predicts a drop-off at low $p_T$.  Data from $\gamma\gamma\rightarrow ee$ in ultra-peripheral collisions also does not show this increase \cite{Adams:2004rz}.  The peak of the pair $p_T$ distribution scales roughly as $\sqrt{1.5M_{ee}}/\gamma$ \cite{Baltz:2009jk}.    In Ref. \cite{Adams:2004rz}, the data diverged from the equivalent photon calculation for $p_T < 20$ MeV/c, for a sample with $M_{ee}> 140$ MeV/c$^2$. If the $\sqrt{M_{ee}}$ scaling holds, the calculated $p_T^2$ spectrum should be OK for $p_T> 35$ MeV/c, or $p_T^2> 0.001$ (GeV/c)$^2$. 

Following the STAR Collaboration \cite{Shuai}, these curves are fit to exponential distributions,
\begin{equation}
\frac{dN}{dp_T^2} = a\exp(-bp_T^2)
\label{eq:expo}
\end{equation}
for events in three mass regions, 0.4 to 0.76 GeV/c$^2$, 0.76 to 1.2 GeV/c$^2$ and 1.2 to 2.6 GeV/c$^2$.  The fit region was chosen to avoid the upturn at low $p_T$:  0.002 (GeV/c)$^2$  $< p_T^2 < $ 0.02 (GeV/c)$^2$.  The fits are shown with the colored lines in the bottom panel of Fig. \ref{fig:pairpt}.  The slopes
are $430 \pm 4$ (GeV/c)$^{-2}$, $402 \pm 6$ (GeV/c)$^{-2}$ and  $367 \pm 9$ (GeV/c)$^{-2}$ for the low, medium and high mass regions respectively.  These numbers are near the upper end of the uncertainty ranges reported by STAR \cite{Shuai}, but the trend with increasing mass agrees well with the data.  It should be noted that STAR used a significantly different fit range, 0.0004 (GeV/c)$^2$  $< p_T^2  <$  0.0064 (GeV/c)$^2$.   That range overlaps with the low-$p_T$ upturn  in Fig. \ref{fig:pairpt} and would have led to significantly larger fit slopes, in disagreement with the STAR data.  Even for the chosen range, the slope depends slightly on the chosen fit range.  

These slope differences are not surprising.  For $k_T \gg k/\gamma$, $dN/dk_T$ from Eq. \ref{eq:photonpt} scales as $F^2(k_T^2)/k_T$.  This naively implies similar slopes, but $k$ scales linearly with pair mass (with some rapidity dependence), so the location of the transition to the $k_T \gg k/\gamma$ varies with pair mass.  The contribution to pair $p_T$ from $\theta$, Eq. 3, should also scale with pair mass, with, for the lepton pseudorapidity cut, a significant dependence on the pair rapidity. 

Similar fits were made to the Au-Au, U-U and Pb-Pb distributions, over the range 400 MeV to 4 GeV$^2$ . They yielded slopes of $404\pm 3$ (GeV/c)$^{-2}$,  $502\pm 4$ (GeV/c)$^{-2}$ and $483\pm 5$ (GeV/c)$^{-2}$, respectively.  The Au-Au and U-U points have a similar trend to the STAR fits, albeit with a larger separation between the two slopes. 

\section{Conclusions}

Two-photon production of lepton pairs in ultra-relativistic collisions is a well-understood process, studied in many ultra-peripheral collision analyses.  In this work, we have studies the two-photon production of lepton pairs in peripheral collisions, and found that it describes the general characteristics of the STAR observations of a continuum excess of $e^+e^-$ pairs at low $p_T$ quite well.  ALICE did not observe this continuum; this is expected because of the required minimum $p_T$ for muons to be observable in their forward muon spectrometer.   However, it should be visible in a future ALICE mid-rapidity study if a low enough lepton $p_T$ cut can be applied. 

I thank Jamie Dunlop, Michael Lomnitz, Lijuan Ruan, Shuai Yang, Wangmei Zha and Zhangbu Xu for useful discussions.
This work was funded by the U.S. Department of Energy under contract number DE-AC-76SF00098.

\end{document}